\documentclass[conference]{IEEEtran}
\IEEEoverridecommandlockouts
\usepackage{cite}
\usepackage{amsmath,amssymb,amsfonts}
\usepackage{algorithmic}
\usepackage{graphicx}
\usepackage{textcomp}
\usepackage{xcolor}
\usepackage{bm}
\usepackage{tcolorbox}

\usepackage{caption}
\usepackage{subcaption}
\usepackage{bigints}
\usepackage{threeparttable}
\allowdisplaybreaks[4]
\def\BibTeX{{\rm B\kern-.05em{\sc i\kern-.025em b}\kern-.08em
    T\kern-.1667em\lower.7ex\hbox{E}\kern-.125emX}}
\newcommand{\tabsname}{{\tt TABS}}

\begin{document}

\title{Trajectory-adaptive Beam Shaping: Towards Beam-Management-Free Near-field Communications 
}
\author{
\IEEEauthorblockN{Sicong Ye\textsuperscript{1}, Yulan Gao\textsuperscript{1}, Ming Xiao\textsuperscript{1}, Peng Wang\textsuperscript{2}, Marios Poulakis\textsuperscript{2}, and Ulrik Imberg\textsuperscript{2}}
\IEEEauthorblockA{\textsuperscript{1}Division of Information Science and Engineering, KTH Royal Institute of Technology, 100 44 Stockholm, Sweden}
\IEEEauthorblockA{\textsuperscript{2}Huawei Technologies Sweden AB, Skalholtsgatan 9-11, 164 40 Kista, Sweden}
\IEEEauthorblockA{Email: \{sicongy, yulang, mingx\}@kth.se, \{peng.wang1, marios.poulakis, ulrik.imberg\}@huawei.com}
}

\maketitle

\begin{abstract}
The quest for higher wireless carrier frequencies spanning the millimeter-wave (mmWave) and Terahertz (THz) bands heralds substantial enhancements in data throughput and spectral efficiency for next-generation wireless networks. 
However, these gains come at the cost of severe path loss and a heightened risk of beam misalignment due to user mobility, especially pronounced in near-field communication. 
Traditional solutions rely on extremely directional beamforming and frequent beam updates via beam management, but such techniques impose formidable computational and signaling overhead. 
In response, we propose a novel approach termed trajectory-adaptive beam shaping (\tabsname{}) that eliminates the need for real-time beam management by shaping the electromagnetic wavefront to follow the user's predefined trajectory. 
Drawing inspiration from self-accelerating beams in optics, \tabsname{} concentrates energy along pre-defined curved paths corresponding to the user's motion without requiring real-time beam reconfiguration. 
We further introduce a dedicated quantitative metric to characterize performance under the \tabsname{} framework. 
Comprehensive simulations substantiate the superiority of \tabsname{} in terms of link performance, overhead reduction, and implementation complexity. 
\end{abstract}

\begin{IEEEkeywords}
Near-field communication, wavefront engineering, beam-management-free, self-accelerating beam, trajectory-adaptive beam shaping.
\end{IEEEkeywords}

\section{Introduction}
Driven by the unprecedented proliferation of wireless devices and escalating demand for ultra-high data rates, the migration toward  higher-frequency bands such as millimeter-wave (mmWave) and even Terahertz (THz) has emerged as an indispensable strategy for realizing next-generation wireless communications \cite{7959169, 9390169}. 
Nevertheless, exploiting these higher frequencies presents considerable propagation hurdles, prominently severe free-space path loss coupled with pronounced atmospheric absorption, inherently constraining the achievable transmission distance \cite{7820226}.
To overcome these limitations and ensure robust communication links, the deployment of extremely large antenna arrays (ELAA) enabling highly directional beamforming has been widely embraced. 
By precisely shaping sharp, narrow electromagnetic beams, these arrays effectively focus radiate power towards specific spatial directions or focal points, thereby significantly augmenting link budgets and markedly improving the receiver's signal-to-noise ratio (SNR) \cite{10734395}.

Moreover, as carrier frequencies continue to increase and antenna array apertures expand accordingly to compensate the increasing signal losses, the communication range transitions from far-field to near-field regions \cite{10734395}, wherein conventional plane-wave channel models no longer suffice. In the near-field regime, the spherical-wave propagation model accurately describes EM wave behavior, enabling precise beam-focusing capabilities that concentrate energy onto specific focal points rather than simple directional steering \cite{10734395}. While this advanced capability has great potential for improving system performance, it simultaneously increases the sensitivity of communication links to receiver mobility or location uncertainty, further exacerbating the challenges associated with beam misalignment. Consequently, beam management is indispensable for maintaining stable, high-quality wireless communication. Beam tracking continuously adapts the beam configuration according to receiver movements and channel variations, ensuring that the transmitter and receiver remain well-aligned. However, existing beam-tracking methods, including those based on Bayesian statistics, and machine learning techniques, still face notable limitations in their computational complexity, and high signaling overhead, especially in highly dynamic environments \cite{yi2024beam}. 

Recently, wavefront engineering has emerged as a powerful approach within the field of optics, enabling comprehensive manipulation of radiated EM waves to generate diverse beams with tailored spatial properties. A prominent example of such engineered wavefront is the Airy beam, which represents the first experimentally demonstrated class of self-accelerating optical beams \cite{chen2012nonlinear, efremidis2019airy}. The Airy beam is particularly distinguished by its unique property of inherent curvature in the main lobe trajectory, facilitating beam propagation along a predefined curved path without external steering. Due to these unique self-accelerating characteristics, Airy beams have received substantial attention and have found numerous potential applications within optical domains \cite{efremidis2019airy}. Despite their notable advantages, these wavefront engineering techniques have predominantly been confined to optical frequencies, with limited exploration within wireless communication frequency bands. It is therefore of paramount importance to bridge this technological gap and extend wavefront engineering methodologies into mmWave and sub-THz wireless communications.

Additionally, in many practical wireless communication scenarios, knowledge of the receiver's trajectory can be effectively acquired or accurately predicted. On the one hand, certain communication scenarios naturally involve deterministic receiver trajectories, such as vehicles moving along highway lanes or high-speed trains traveling along fixed railway tracks. On the other hand, even in dynamic settings, future trajectories can often be accurately predicted within short periods, as exemplified by robotics in industrial Internet-of-Things (IoT) applications or automated vehicles in structured environments. In these cases, despite potential variations in speed or minor uncertainties in location, trajectory information can be leveraged proactively to enhance beam management.

Motivated by the distinctive self-accelerating properties of optical wavefront-engineered beams and the practical availability of receiver trajectory information in wireless communications, this paper is, to the best of our knowledge, the first to apply such beam design principles to near-field beam management, thereby enabling beam-management-free operation. We introduce trajectory-adaptive beam shaping (\tabsname{}), a framework that systematically shapes near-field beams to intrinsically cover predefined user trajectories, substantially reducing or even eliminating the need for conventional beam-management procedures.
Additionally, we propose a new performance metric to rigorously quantify the effectiveness of \tabsname{} in maintaining high-quality communication along user trajectories. Through comprehensive numerical simulations and comparative analyses, we demonstrate that the \tabsname{} approach substantially outperforms conventional beam focusing schemes, achieving superior link robustness while simultaneously reducing signaling overhead and computational complexity.

\section{System Model and Performance Metric}
\subsection{System Model}
\begin{figure}[!t]
    \centering
\includegraphics[width=0.6\linewidth]{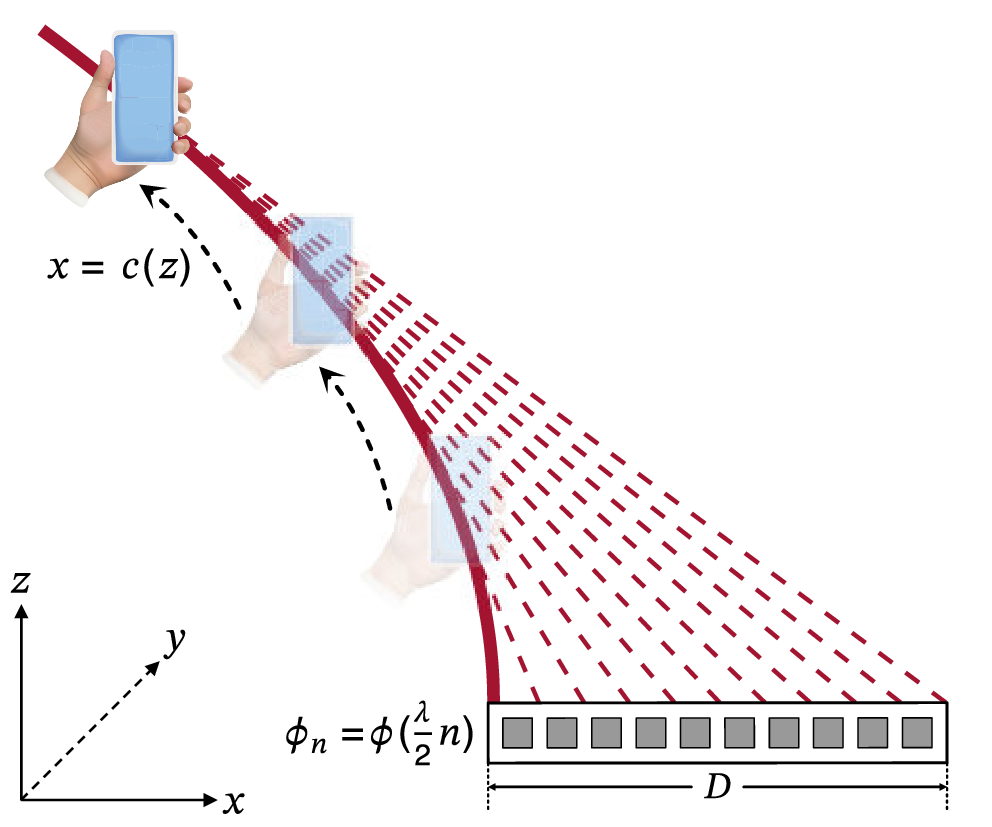}
\caption{Illustration of the considered near-field communication scenario with \tabsname{}.}
    \label{fig:sm}
\end{figure}

We consider a downlink near-field communication system as depicted in Fig.~\ref{fig:sm}, where a transmitter equipped with a $N$-element half-wavelength spaced uniform linear antenna array (ULA) communicates with a single-antenna mobile receiver. Specifically, we establish a Cartesian coordinate system in which the transmitter ULA is placed along the $x$-axis, spanning coordinates from $0$ to $D$, where $D=(N-1)\lambda/2$ is the aperture length of the ULA and $\lambda$ is the signal wavelength. 
A single-antenna receiver moves along a predefined trajectory segment within the near-field region in $x-z$ plane\footnote{The beam design and propagation analysis along the $y$-axis follow analogously to those along the $x$-axis. Therefore, for clarity and without loss of generality, we restrict our analysis to the two-dimensional $x-z$ plane.}, defined by $\mathcal{S} = \{(x,z)\,|\, x = c(z), z \in [z_{\text{start}}, z_{\text{end}}]\}$, where the precise instantaneous position along this trajectory is considered unknown at the transmitter side during transmission. 
We consider an analog beamforming architecture, where each antenna element transmits with constant amplitude. Consequently, the corresponding analog beamforming vector is given by: 
\begin{equation} 
\bm \omega = \frac{1}{\sqrt{N}} \left[e^{j\phi_1}, e^{j\phi_2}, \ldots, e^{j\phi_N}\right]^T, 
\end{equation}
where $\phi_n$ denotes the phase shift applied to the $n$-th antenna.

In near-field scenarios within the Fresnel region \cite{10734395}, each transmitter-receiver antenna pair experiences a distinct propagation path. Unlike far field approximation, both amplitude and phase variations across the array become significant and thus cannot be neglected \cite{10734395}. Therefore, in order to accurately capture the characteristics of near-field propagation, the line-of-sight (LoS) near-field channel for the $n$-th antenna under the geometric free space propagation assumption \cite{10734395} is modeled as
\begin{align}\label{eqn:nf-channel}
    h_n(x,z) = \frac{1}{r_n(x,z)} e^{-j k_0  r_n(x,z)},
\end{align}
where $r_n$ denotes the distance of the $n$-th antenna of transmitter to the receiver and $k_0 = 2\pi/\lambda$ is the wave number. Consequently, the near-field channel between the transmitter and receiver can be expressed by stacking the channel coefficients of all $N$ antenna elements into a vector: $\bm h = [h_1(x,z), h_2(x,z), \cdots, h_N(x,z)]^T$.


Let $s \in \mathcal{C}$ denote the transmitted signal with unit power, i.e., $\mathbb{E}[s s^H] = 1$. The noiseless received signal at the receiver side can be expressed as
\begin{equation}\label{eqn:received}
    y_{\text{rx}}^{\text{noiseless}}(x,z) = {\bm h(x,z)}^H {\bm \omega} s,
\end{equation}
As a result, the corresponding signal strength is given by
\begin{equation}
    I(x,z) = |{\bm h(x,z)}^H {\bm \omega}|.
\end{equation}
The design objective is to engineer the transmit wavefront so that the radiated energy consistently follows the predefined trajectory, ensuring robust communication without requiring real-time beam management.


\subsection{Performance Metric}
To quantitatively evaluate the ability of \tabsname{} to sustain the quality of service (QoS) provided by \tabsname{} along the trajectory of the moving receiver compared with conventional beam management method, we introduce the spatial outage reliability, denoted spatial outage reliability, $R_{\mathcal{S}}(\gamma)$. This metric measures the fraction of the total trajectory $\mathcal{S}$ over which a chosen performance indicator exceeds a predefined threshold $\gamma$. The threshold $\gamma$ can represent any relevant communication quality measure, such as received signal strength, achievable throughput, or packet error rate.

In this paper, we adopt received signal strength as the primary performance indicator, given its direct relevance to quality of communication. Assuming the receiver moves along a predefined trajectory $\mathcal{S}$. The spatial outage reliability is then expressed as
\begin{equation}
    R_{\mathcal{S}} (\gamma) = \frac{\mu(\{(x,z) \, | \, I\left( c(z),z \right)\ge\gamma, \,z \in [z_{\text{start}}, z_{\text{end}}]\})}{\mu(\mathcal{S})}, 
\end{equation}
where $\mu(\cdot)$ denotes the Lebesgue measure, corresponding to the geometric length of the set. A higher value of $R_{\mathcal{S}} (\gamma)$ indicates that a greater proportion of the trajectory enjoys adequate signal quality, reflecting both the spatial robustness and reliability of the beam design.




\section{Methodology of Trajectory Design}
\subsection{Wavefront Engineering}

We provide a detailed derivation of the general wavefront engineering methodology employed for designing trajectory-based beams in near-field wireless communication scenarios. Specifically, we start from fundamental electromagnetic wave propagation principles and systematically derive the relationship between a predefined beam trajectory and the required initial phase distribution at the transmitting antenna aperture.

We assume the phase shift on the $n$-th antenna element $\phi_n$ is obtained by sampling from continuous phase function $\phi(\xi), \xi\in[0,D]$, which is given by
\begin{equation}\label{eqn:dis} \phi_n=\phi\left((n-1)\frac{\lambda}{2}\right),\quad n=1,2,\dots,N. \end{equation}
Consequently, the received signal electromagnetic field at a spatial location $(x,z)$ in \eqref{eqn:received} can be rewritten as
\begin{equation}
    \begin{split}
        y_{\text{rx}}^{\text{noiseless}}(x,z) & = \int_0^D h(x,z;\xi)^* e^{j\phi(\xi)} \, d\xi \\
        & = \int_0^D \frac{1}{\sqrt{(x-\xi)^2 + z^2}} e^{\phi(\xi) + k_0 \sqrt{(x-\xi)^2 + z^2}} d\xi \\
        & = \int_0^D \frac{1}{\sqrt{(x-\xi)^2 + z^2}} e^{\Psi(\xi; x,z)} d\xi,
    \end{split}
\end{equation}
where $\Psi(\xi; x,z)=\phi(\xi) + k_0 \sqrt{(x-\xi)^2 + z^2}$ is the total phase accumulated by the electromagnetic field at a spatial location $(x,z)$ and phase at aperture $\phi(\xi)$.


To simplify subsequent analysis, we adopt the Fresnel approximation by transforming spherical wavefront to parabolic wavefront \cite{goodman2005introduction}, yielding:  
\begin{equation}
    \Psi(\xi; x,z)  \stackrel{(a)}{\approx} \phi(\xi) + k_0 \left( z + \frac{(x-\xi)^2}{2z} \right),
\end{equation}
where $(a)$ is based on the Taylor expansion $\sqrt{1+x} \approx 1 + \frac{1}{2}x$ for $|x|<1$. Applying first and second order stationary-phase condition from catastrophe theory \cite{berry1980iv}, i.e., $\frac{\partial \Psi(\xi; x,z)}{\partial \xi} = 0$ and $\frac{\partial^2 \Psi(\xi; x,z)}{\partial \xi^2} = 0$, we have
\begin{align}
    & \frac{d\phi(\xi)}{d\xi} = (x - \xi)\frac{k_0}{z}, \label{eqn:ray} \\
    & \frac{d^2\phi(\xi)}{d\xi^2} = -\frac{k_0}{z}.
\end{align}
To ensure a globally maximal field intensity along the desired trajectory, the relationship between trajectory coordinates and aperture coordinates must be smooth and single-valued, i.e., $\xi(x)$. Thus, differentiating \eqref{eqn:ray} with respect to the spatial coordinate $x$, we obtain: \begin{equation}\label{eqn:ddphidx1} 
\frac{d}{d x}  \frac{d\phi(\xi)}{d\xi} = \frac{d^2\phi(\xi)}{d\xi^2} \frac{d\xi}{dx} = -\frac{k_0}{z} \frac{d\xi}{dx}=-\frac{k_0}{g(x)} \frac{d\xi}{dx}, 
\end{equation} 
where we define the inverse trajectory function $g(x) = c^{-1}(x) = z$. Additionally, directly substituting $z = g(x)$ into \eqref{eqn:ray} and differentiating again yields: \begin{equation}\label{eqn:ddphidx2} 
\frac{d}{d x} \frac{\phi(\xi)}{d\xi} = k_0 \frac{\left(1-d\xi/dx\right)g(x)-g(x)/dx\left(x-\xi\right)}{g^2(x)}. 
\end{equation} 
By equating \eqref{eqn:ddphidx1} and \eqref{eqn:ddphidx2}, we derive the trajectory-envelope condition explicitly as: 
\begin{equation}\label{eqn:tan ray} 
x= \xi + \frac{g(x)}{dg(x)/dx} = \xi+\frac{dc(z)}{dz}z = \tan{\theta} \,z, 
\end{equation} where $\theta$ is the deviation angle from $z$-axis. This relation signifies that each ray emitted from the aperture is tangent to the trajectory $c(z)$, i.e., $\xi(x) = x - z \frac{dc(z)}{dz}$. Therefore, the designed trajectory acts as the envelope of all rays emerging from the aperture and is thus recognized as a caustic\footnote{In geometric optics~\cite{lynch2001color}, a caustic refers to the envelope formed by rays reflected or refracted by a curved manifold.}.

On the other hand, according to the generalized Snell’s law \cite{yu2011light}, the imposed aperture phase gradient dictates the local deviation angle $\theta$ of the normally incident beam through the relationship: \begin{equation} d\phi(\xi)/d\xi=k_0\sin\theta. \end{equation} Utilizing standard trigonometric identities, the required aperture-plane phase gradient for achieving the predefined beam trajectory can be explicitly derived as: \begin{equation}\label{eqn:phase_numerical} \phi(\xi)= \int_0^\xi k_0\frac{dc(z)/dz}{\sqrt{1+\left(dc(z)/dz\right)^2}}d\xi. \end{equation} 
Consequently, the corresponding phase on the $n$-th antenna across the array is given by  \eqref{eqn:dis}. While the phase profile for arbitrary beam trajectories typically requires numerical evaluation, analytical closed-form solutions exist for specific classes of convex trajectories, such as circular or parabolic curves~\cite{froehly2011arbitrary,penciu2015closed}.

\subsection{Case Demonstrations}
To demonstrate the practical applicability of the phase design methodology\footnote{According to \eqref{eqn:phase_numerical}, beam trajectory is fully characterized by the aperture-plane phase profile. Consequently, this paper focuses exclusively on the design of the initial phase distribution, assuming a constant amplitude profile, i.e., $\psi_0(\xi) = e^{j\phi(\xi)}$. The investigation of amplitude modulation effects is beyond the scope of this paper and will be addressed in future work.} described previously, two representative accelerating beam trajectories, circular and parabolic, are presented and analyzed because of their analytical tractability.

Consider first an accelerating beam trajectory defined by a circular caustic described as:
\begin{equation}
    x = c(z) = \sqrt{R^2 - z^2},
\end{equation}
where $R>0$ denotes the radius of the circular path. Although the initial phase distribution required to realize an arbitrary trajectory typically involves numerical computation based on \eqref{eqn:phase_numerical}, an analytical closed-form solution for a circular trajectory has been established in the literature \cite{penciu2015closed}. The initial aperture-plane phase profile is explicitly given by:
\begin{equation}
    \phi(\xi) = k_0 R \left( \sqrt{({\xi}/{R})^2 - 1} - \sec^{-1} \left( {\xi}/{R} \right) \right).
\end{equation}

The second considered case is a parabolic trajectory, which is widely recognized due to its connection to Airy-type accelerating beams. This trajectory can be expressed as:
\begin{equation}
    x = \alpha z^2,
\end{equation}
where the curvature parameter $\alpha> 0$ controls the rate of acceleration along the trajectory. Through appropriate derivations, the corresponding input-plane phase profile required to generate this parabolic beam trajectory can be expressed analytically as:
\begin{equation}
    \phi(\xi) = - \frac{4 \alpha k_0 \xi}{3} \sqrt{\frac{\xi}{\alpha}} \, {}_2F_1 \left( \frac{1}{2}, \frac{3}{2}; \frac{5}{2}; -4 \alpha \xi \right),
\end{equation}
where ${}_2F_1(\cdot)$ is the hypergeometric function \cite{aomoto2011theory}.

\begin{figure}[t]
     \centering
     \hfill
     \begin{subfigure}[b]{0.24\textwidth}
         \centering
         \includegraphics[width=0.9\textwidth]{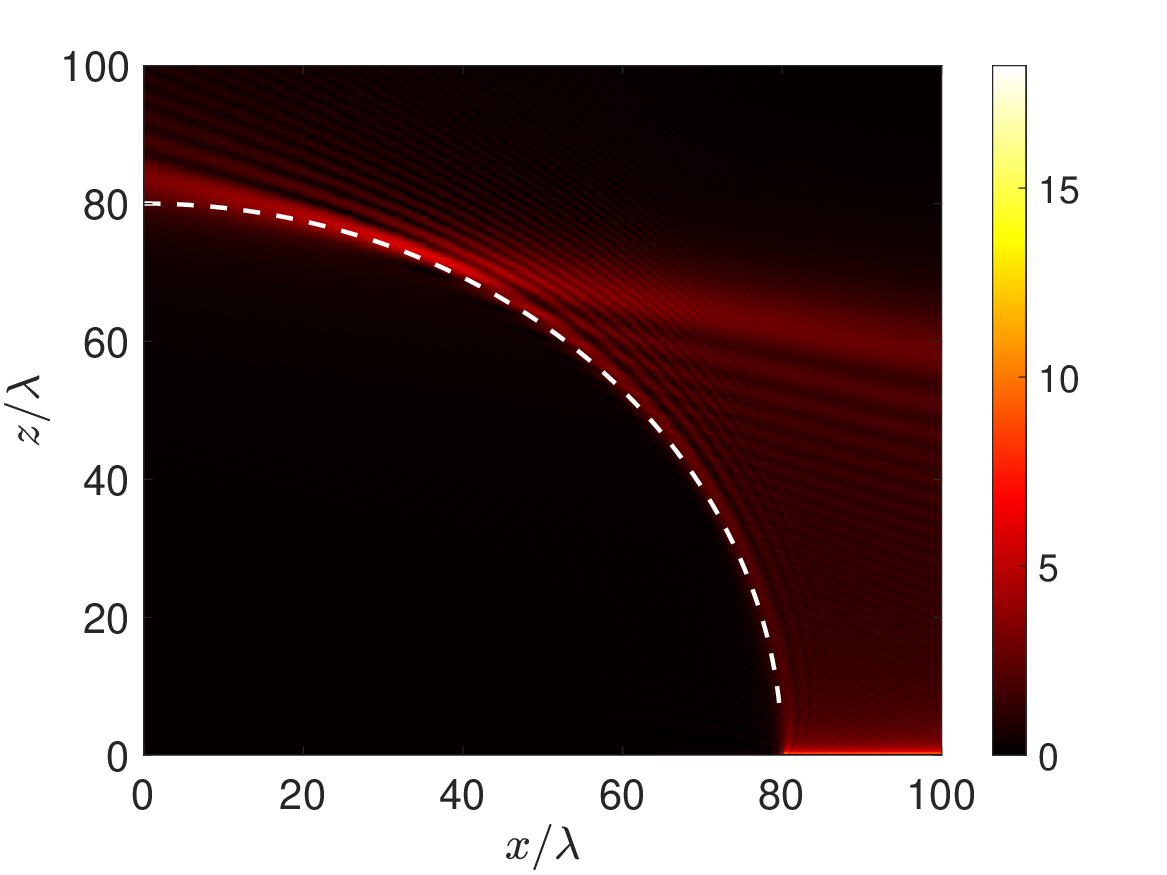}
         \caption{Circular trajectory with radius $R = 80$.}
         \label{fig:examples_circle_80R}
     \end{subfigure}
     \begin{subfigure}[b]{0.24\textwidth}
         \centering
         \includegraphics[width=0.9\textwidth]{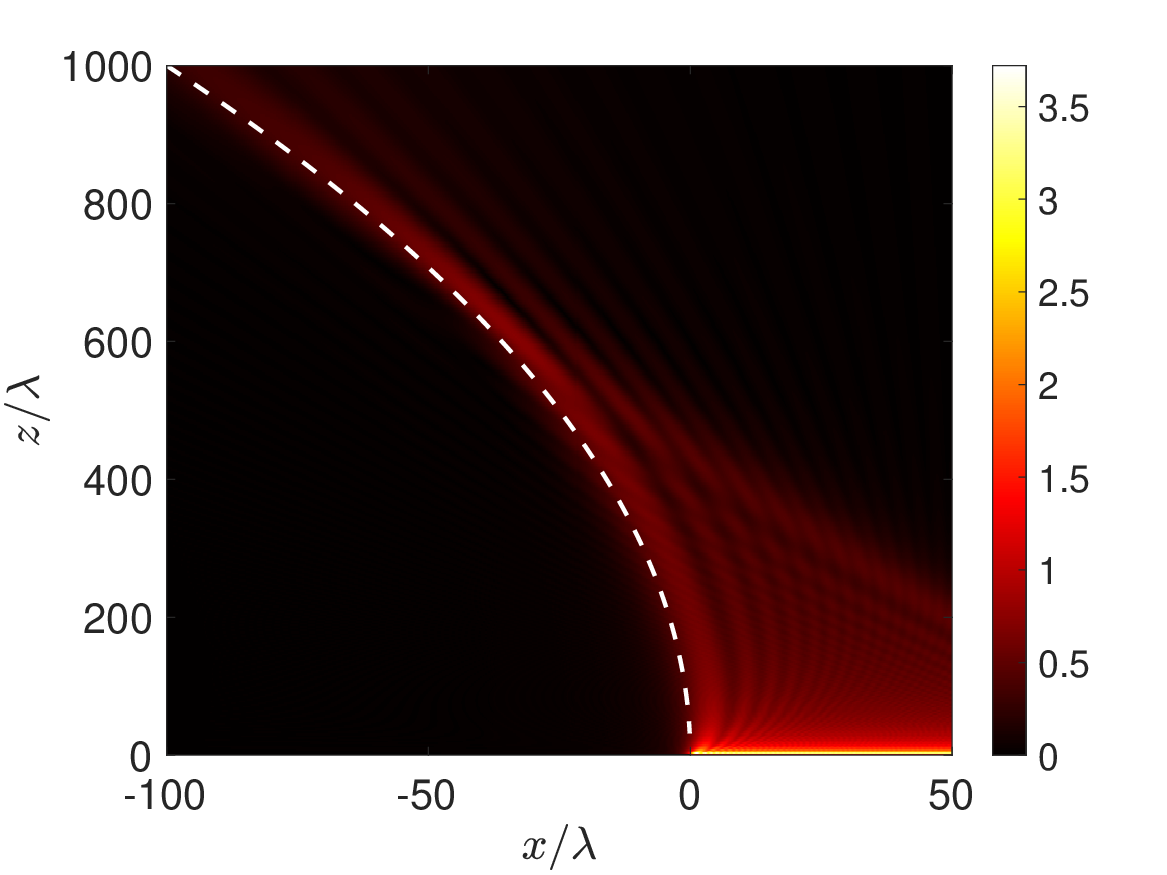}
         \caption{Parabolic trajectory with curvature parameter $\alpha=0.0001$.}
         \label{fig:examples_parabolic_1a}
     \end{subfigure}
        \caption{Intensity patterns numerical examples of accelerating beam trajectories designed through phase control at the input aperture.}
        \label{fig:examples}
\end{figure}

To validate the presented analytical solutions and illustrate beam propagation along the designed trajectories, numerical simulations were performed, as depicted in Fig. \ref{fig:examples}. Fig. 
\ref{fig:examples_circle_80R} 
shows the accelerating beams generated along circular trajectories with radius $R=80$. The simulated beam intensity patterns clearly exhibit excellent agreement with the analytically defined circular paths (white dashed lines), thus verifying the accuracy of the closed-form phase solutions. Fig. 
\ref{fig:examples_parabolic_1a} 
demonstrates beams following parabolic trajectories with curvature parameters $\alpha=0.0001$. In both cases, the simulated intensity distributions precisely follow the theoretical parabolic curves, confirming the correctness of the derived analytical phase expressions. A decrease in the curvature parameter $\alpha$ results in a slower curvature and more gradual bending trajectory, indicating the flexibility offered by this approach in adapting beam propagation paths to diverse near-field communication scenarios.

Overall, these results validate the effectiveness and flexibility of the phase design in accurately controlling beam trajectories for near-field communication.

\addtolength{\topmargin}{0.03in}

\subsection{Computational Complexity Analysis}
Computational complexity critically influences the practical deployment of beamforming techniques in dynamic communication scenarios. Conventional beam-tracking methods, such as Kalman filters (KF) \cite{8851228}, particle filters (PF) \cite{9189808}, and auxiliary beam pair (ABP) algorithms \cite{7929295}, rely on continuous iterative operations, real-time estimation, and sequential updates. Consequently, these conventional approaches consistently incur substantial computational overhead due to repeated state estimations, even when the user trajectory is predetermined or known in advance.

Conversely, the proposed \tabsname{} employs a fundamentally simpler computational paradigm. It requires only a single initial phase profile computation, eliminating real-time iterative processing. Table~\ref{tab:complexity_comparison} summarizes the computational complexity comparison, clearly highlighting the efficiency advantage of \tabsname{}.

\begin{table} 
\centering 
\begin{threeparttable}
\caption{Computational Complexity Comparison.} \label{tab:complexity_comparison}
\begin{tabular}{|c|c|c|} 
\hline 
\textbf{Beam Tracking Method} & \textbf{Computational Complexity} & \textbf{Reference } \\ 
\hline \hline KF & $\mathcal{O}(N_i n_x^3)+\mathcal{O}({N})$ & \cite{8851228}\\ 
\hline PF & $\mathcal{O}(N_p N_i n_x^2)+\mathcal{O}({N})$ & \cite{9189808} \\ 
\hline ABP & $\mathcal{O}(N_a N_i n_x^2)+\mathcal{O}({N})$ & \cite{7929295} \\ 
\hline \textbf{\tabsname{}} & $\mathcal{O}({N})$ & -- \\ 
\hline 
\end{tabular} 
\begin{tablenotes}
\item{\bf\em Notations:} $N_i$: number of iterations; $n_x$: dimension of state-space; $N_p$: number of particles in PF; $N_a$: number of auxiliary beam pairs.
\end{tablenotes}
\end{threeparttable}
\end{table}


\subsection{Signaling Overhead Analysis}
Beyond its computational simplicity, \tabsname{} significantly reduces signaling overhead compared to conventional beam management methods specified by 3GPP New Radio (NR) standards for mmWave and sub-THz bands. Traditional beam management primarily incurs overhead due to repeated synchronization signal (SS) bursts, CSI-RS transmissions, and beam reporting processes necessary for maintaining beam alignment \cite{3gpp}. The dominant overhead arises from periodic SS bursts used in beam sweeping, consuming considerable resources—up to 43\% at short burst intervals \cite{giordani2018tutorial}.

In contrast, \tabsname{} eliminates periodic beam-sweeping and beam refinement at transmit- and receive-end procedure after initial access, as trajectory information enables beam shaping along the user's predicted path without ongoing directional scans. Consequently, \tabsname{} substantially reduces resource consumption and overall signaling complexity.

\section{Simulation Results and Analysis}
We conduct a comprehensive numerical evaluation to illustrate the advantages offered by \tabsname{} in near-field communication scenarios, particularly emphasizing their capability to eliminate the necessity of conventional beam management procedures. We first describe the detailed simulation setup and parameters employed in the evaluation, clearly defining the system configuration and the performance metrics adopted for comparative analyses. Subsequently, extensive simulation results are provided, illustrating the superior communication performance, significant reduction in beam management overhead, and reduced computational complexity achieved by \tabsname{} relative to traditional beam-management techniques. Collectively, these results substantiate the effectiveness, practical feasibility, and performance advantages of utilizing trajectory-based beams for robust and efficient near-field communication systems without requiring continuous beam alignment.


\subsection{Simulation Setup}
We present the simulation parameters and assumptions adopted to rigorously evaluate the effectiveness of \tabsname{} compared with conventional beam-management methods. We consider a near-field communication scenario consisting of a transmitter equipped with a ULA characterized by an aperture length $D=500\lambda$. Accordingly, this ULA consists of $N = 1001$ antenna elements, each spaced at half the carrier wavelength\footnote{While the number of antenna elements may appear large, this array configuration corresponds to realistic physical dimensions in current and forthcoming wireless communication systems. For instance, this array aperture ranges from approximately $2$ to $6$ meters within the 5G frequency range 2 (FR2) operating at millimeter-wave frequencies \cite{7959169}.}. The single-antenna receiver is assumed to move along a predefined parabolic trajectory given by $x = c(z) = \alpha z^2$, with curvature parameter $\alpha = 0.00025$.

The employed channel model assumes a LoS near-field propagation environment, described by \eqref{eqn:nf-channel}. To evaluate the effectiveness of the proposed TABS method, we compare it against three baseline schemes:
\begin{itemize}
    \item Conventional BF: A conventional beam focusing approach targeting a fixed point on the trajectory, without any adaptation.
    \item Multi-Point BF: To improve spatial coverage, this scheme forms multiple focused beams along several points on the trajectory and linearly superimposes them with equal power allocation.
    \item Multi-BF: This method assumes that the receiver’s instantaneous location is tracked in real time, and the transmit beam is continuously re-steered toward the updated receiver position. While highly effective in theory, it incurs significant beam management overhead and computational complexity.
\end{itemize}

\subsection{Performance Evaluation}
We conduct a detailed numerical performance evaluation comparing the proposed \tabsname{} method with the conventional beam focusing (BF) approach in near-field communication scenarios. Specifically, we utilize the previously defined spatial outage reliability $R_\mathcal{S}(\gamma)$ to quantify the performance benefits provided by \tabsname{} in beam management. 

\begin{figure}[t]
     \centering
     \begin{subfigure}[b]{0.15\textwidth}
         \centering
         \includegraphics[width=\textwidth]{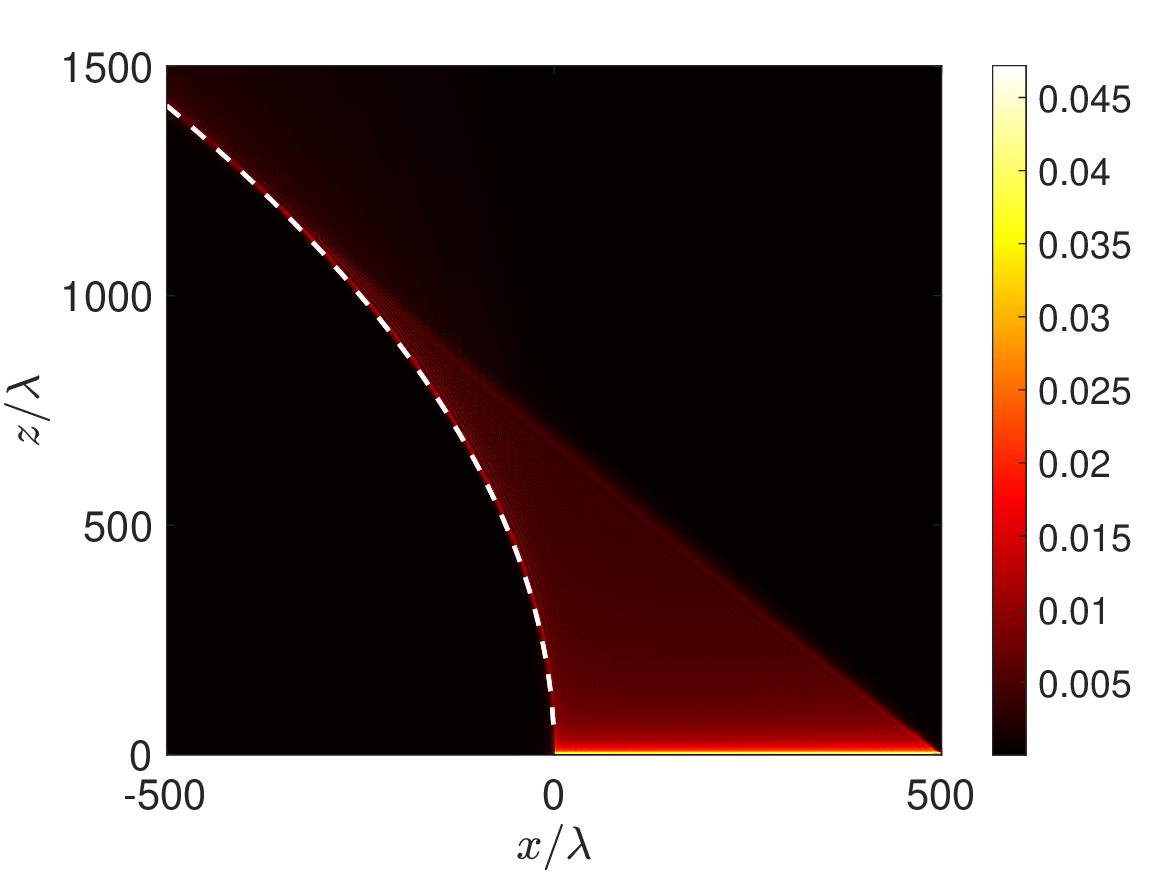}
         \caption{}
     \end{subfigure}
     \hfill
     \begin{subfigure}[b]{0.15\textwidth}
         \centering
         \includegraphics[width=\textwidth]{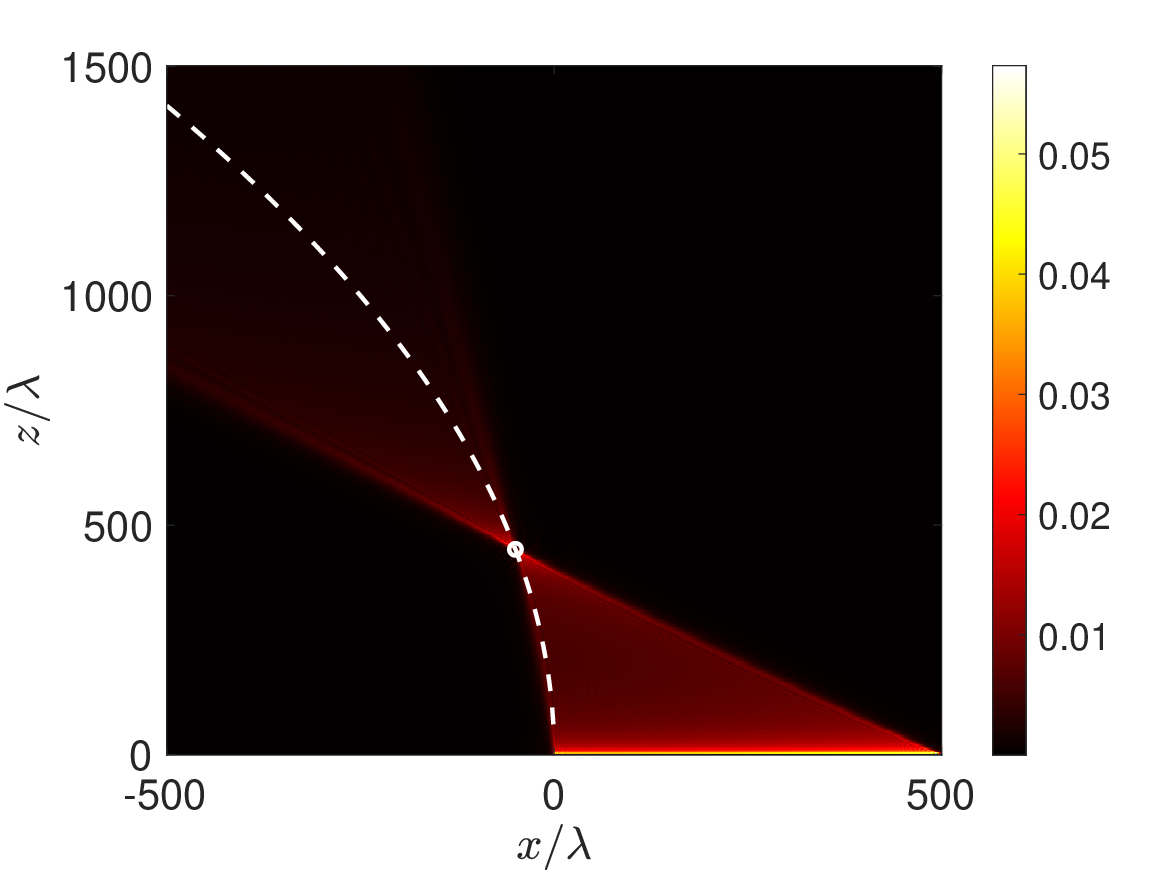}
         \caption{}
     \end{subfigure}
     \hfill
     \begin{subfigure}[b]{0.15\textwidth}
         \centering
         \includegraphics[width=\textwidth]{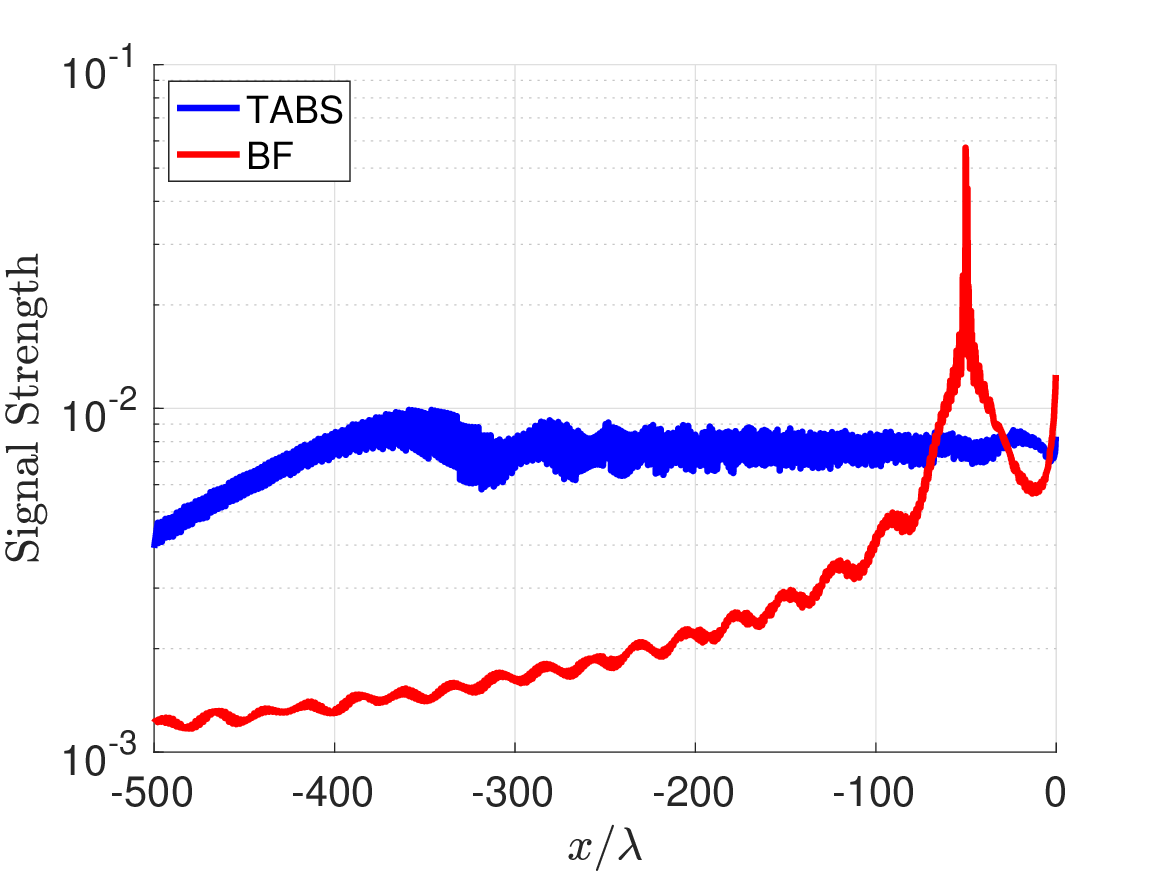}
         \caption{}
     \end{subfigure}
        \caption{Beam propagation patterns under near-field propagation for (a) \tabsname{} and (b) BF approaches. The dashed white curve denotes the intended receiver trajectory $x = c(z) = \alpha z^2$. (c) Signal strength along the trajectory.}
        \label{fig:beam_prop}
\end{figure}

To illustrate the propagation behavior of \tabsname{} and conventional BF, Fig.~\ref{fig:beam_prop} presents the simulated near-field beam intensity distributions and corresponding signal strength profiles along the intended receiver trajectory. \tabsname{} effectively shapes the wavefront to concentrate energy along the entire parabolic trajectory (dashed white curve), thereby ensuring continuous high-intensity coverage without requiring real-time beam adjustments. In contrast, the BF method can only focuses energy at a single point, resulting in rapid signal decay as the receiver moves away from this point. These results demonstrate that TABS inherently supports beam-management-free operation, while BF requires frequent beam updates to maintain adequate signal strength.

\begin{figure*}[t]
     \centering
     \begin{subfigure}[b]{0.3\textwidth}
        \centering
        \includegraphics[width=0.9\linewidth]{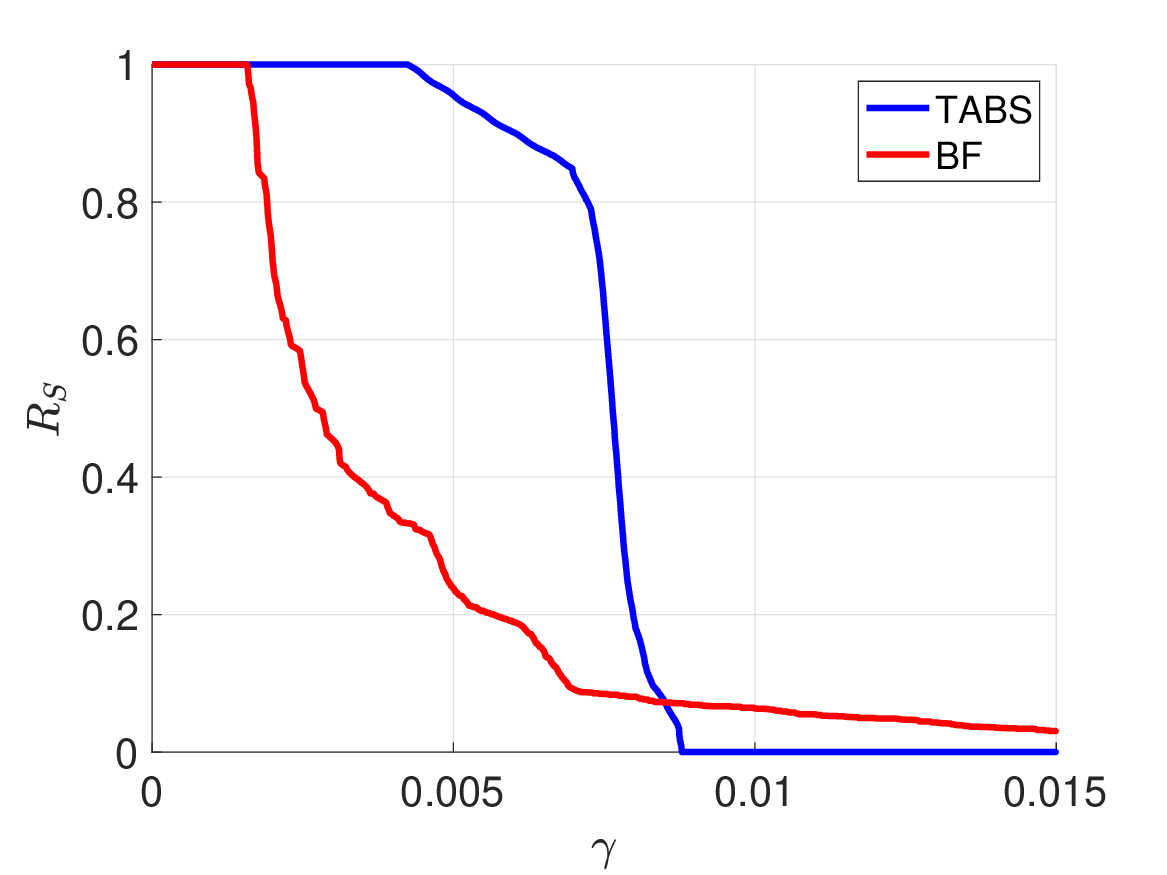}
        \caption{}
        \label{fig:L_gamma}
     \end{subfigure}
     \hfill
     \begin{subfigure}[b]{0.3\textwidth}
        \centering
        \includegraphics[width=0.9\linewidth]{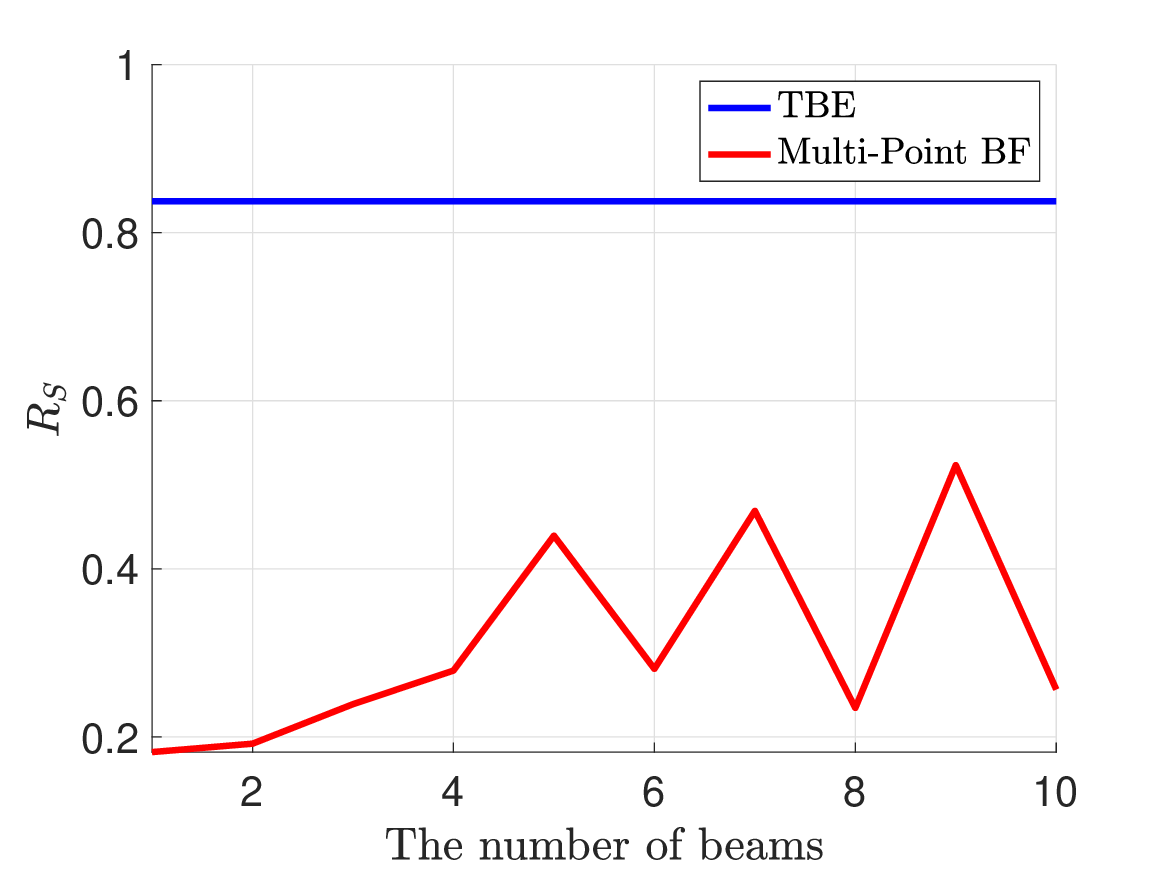}
        \caption{}
        \label{fig:multi-beam}
     \end{subfigure}
     \hfill
     \begin{subfigure}[b]{0.3\textwidth}
         \centering
        \includegraphics[width=0.9\linewidth]{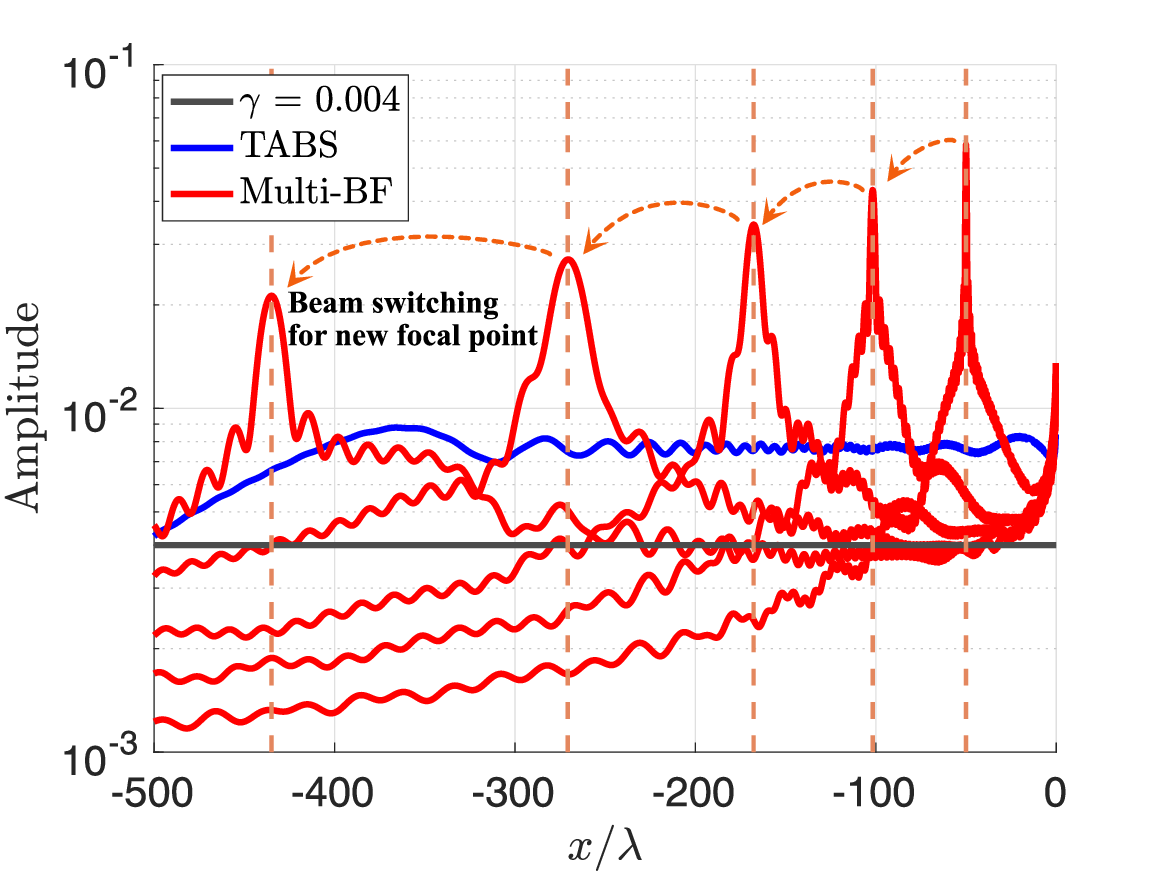}
        \caption{}
        \label{fig:multi-bf}
     \end{subfigure}
    \caption{(a) Comparison of $R_{\mathcal{S}}$ between \tabsname{} and BF schemes vs $\gamma$. (b) $R_{\mathcal{S}}$ versus the number of beams for conventional BF and proposed \tabsname{} methods. (c) Comparison of received signal strength along the trajectory.}
\end{figure*}

We firstly quantitatively evaluate and compare the performance achieved by \tabsname{} and BF approaches at fixed focal positing at $x=-100$ along the trajectory, using the $R_{\mathcal{S}}$ metric across different values of the signal strength threshold $\gamma$, as shown in Fig.~\ref{fig:L_gamma}. It is evident from the results that \tabsname{} consistently outperforms BF, maintaining a higher spatial outage reliability level over a broad range of thresholds. Specifically, at lower threshold values (e.g., $\gamma\leq0.0015$), both \tabsname{} and BF schemes can reliably satisfy the signal strength requirement. However, as the threshold increases to $\gamma=0.005$, the spatial outage reliability achieved by BF dramatically decreases to approximately $20\%$, while the proposed \tabsname{} approach still successfully maintains a near-perfect level. Upon further increasing the threshold to $\gamma=0.01$, the $R_{\mathcal{S}}$ for both methods deteriorates significantly, falling below $5\%$, indicating a failure to consistently maintain adequate communication performance along the trajectory. This comparison clearly highlights the enhanced robustness and improved signal quality provided by \tabsname{}, especially at low-to-moderate signal-strength thresholds.

We further investigate the effectiveness of the proposed \tabsname{} approach compared to conventional BF through simulations involving multiple focal points. Specifically, we uniformly distribute multiple focal points along the receiver's trajectory on the $x$-axis, allocating equal transmission power to each focal point. The resulting performance in terms of the spatial outage reliability with a signal strength threshold $\gamma = 0.007$ illustrated in Fig.~\ref{fig:multi-beam}. As observed, when the focal points increases from one to five, the $R_{\mathcal{S}}$ achieved by the BF method steadily improves. This improvement occurs because increasing the number of focal points enhances the spatial coverage, allowing more trajectory segments to meet the defined signal strength threshold. However, further increasing the number of beams beyond five causes noticeable fluctuations, primarily due to the power dilution effect. This phenomenon arises because the available power is uniformly distributed among a larger number of beams, thereby reducing the individual beam strength and causing the energy at some focal points to fall below the required threshold. In contrast, the proposed \tabsname{} maintains a consistently high $R_{\mathcal{S}}$, significantly outperforming BF. \tabsname{} effectively shapes the transmitted beam along the entire trajectory through a single optimized wavefront design, efficiently concentrating energy where required.

Finally, to demonstrate the inherent advantage of \tabsname{} for scenarios in which the receiver trajectory is known, we present the amplitude distribution of received signals along the trajectory in Fig.~\ref{fig:multi-bf}. We assume here that the exact receiver position is perfectly known at the transmitter for BF scheme. Nevertheless, when employing the conventional BF approach, it is necessary to repeatedly generate new focal points along the trajectory, known as beam switching/recovery. Specifically, each time the receiver moves beyond the trajectory segment where the received signal strength falls below the desired threshold $\gamma$, a new beam-focusing operation must be performed. This requirement introduces considerable signaling overhead and complexity. Conversely, the \tabsname{} approach achieves continuous and consistent signal coverage across the entire trajectory with only a single initial phase-engineering step, thus highlighting significant advantages in terms of simplicity, robustness, and overhead reduction relative to conventional BF methods.



\section{Conclusion}
This paper introduces \tabsname{} as a novel approach to enable beam-management-free communication in near-field wireless systems. By leveraging wavefront engineering, \tabsname{} generates beams that follow predefined user trajectories through a one-time phase design at the transmitter. Compared to conventional beam management methods, \tabsname{} significantly reduces signaling overhead and computational complexity, while maintaining robust link performance. Numerical results confirm that \tabsname{} offers a practical and efficient solution for future near-field networks with predictable user mobility.

\bibliographystyle{ieeetr}
\bibliography{ref}

\end{document}